\documentclass[runningheads]{llncs}
\usepackage[misc]{ifsym}
\usepackage{graphicx}
\usepackage{cite}
\usepackage{amsmath,amssymb,amsfonts}
\usepackage{algorithmic}
\usepackage{graphicx}
\usepackage{float} 
\usepackage{subfigure}
\usepackage{textcomp}
\usepackage{xcolor}
\usepackage{booktabs}
\usepackage{caption}
\usepackage{url}
\usepackage{makecell}
\usepackage{listings}
\usepackage{color} 
\usepackage{multirow}
\usepackage{algorithm}
\usepackage{algorithmic}
\usepackage{CJKutf8}
\usepackage{hyperref}
\usepackage{bbding}
\usepackage{caption,subcaption}
\hypersetup{hypertex=true,
            colorlinks=true,
            linkcolor=blue,
            anchorcolor=blue,
            citecolor=blue}

\begin{document}
\title{UMAIR-FPS: User-aware Multi-modal Animation Illustration Recommendation Fusion with Painting Style\thanks{\small This work was supported in part by the National Natural Science Foundation of China, Grant No. 62366060, 61762092, Open Foundation of Yunnan Key Laboratory of Software Engineering under Grant No.2023SE203.}}
\titlerunning{UMAIR-FPS}
%

\author{Yan Kang\inst{1,2} \and
Hao Lin\inst{1} \and
Mingjian Yang\inst{1(}\Envelope\inst{)}\and
Shin-Jye Lee\inst{3}
}
\authorrunning{Y. Kang et al.}
%
\institute{National Pilot School of Software, Yunnan University, Kunming 650106, China
\email{kangyan@ynu.edu.cn, \{oysterqaq,ymj123\}@mail.ynu.edu.cn}
\and
Yunnan Key Laboratory of Software Engineering, Kunming 650106, China\\
 \and
Institute of Management of Technology, National Yang Ming Chiao Tung University, Hsinchu, Taiwan\\
\email{camhero@gmail.com}}
\maketitle              
\begin{abstract}

The rapid advancement of high-quality image generation models based on AI has generated a deluge of anime illustrations. Recommending illustrations to users has become a challenge. However, existing anime recommendation systems (RS) have focused on text features but still need to integrate image features. In addition, most multi-modal (MM) RS research is constrained by tightly coupled datasets, limiting its applicability to illustrations RS. We propose the \textbf{U}ser-aware \textbf{M}ulti-modal \textbf{A}nimation \textbf{I}llustration \textbf{R}ecommendation \textbf{F}usion with \textbf{P}ainting \textbf{S}tyle (UMAIR-FPS) to tackle these gaps. In the feature extract phase, for image features, we are the first to combine painting style with semantic features to construct a dual-output image encoder for enhancing representation. For text features, we obtain embeddings based on fine-tuning Sentence-Transformers by incorporating domain knowledge that composes a variety of anime text pairs from multilingual mappings, entity relationships, and term explanation perspectives, respectively. In the MM fusion phase, we novelly propose a user-aware multi-modal contribution measurement mechanism to weight MM features dynamically according to user features at the interaction level and employ the DCN-V2 module to model bounded-degree MM crosses effectively. UMAIR-FPS surpasses the SOTA baselines on large real-world datasets.

\end{abstract}
\section{Introduction}
\begin{figure}
 \vspace{-12pt}
\centering  
\subfigure[]{
\label{illusts_with_diff_style}
\includegraphics[width=0.46\textwidth]{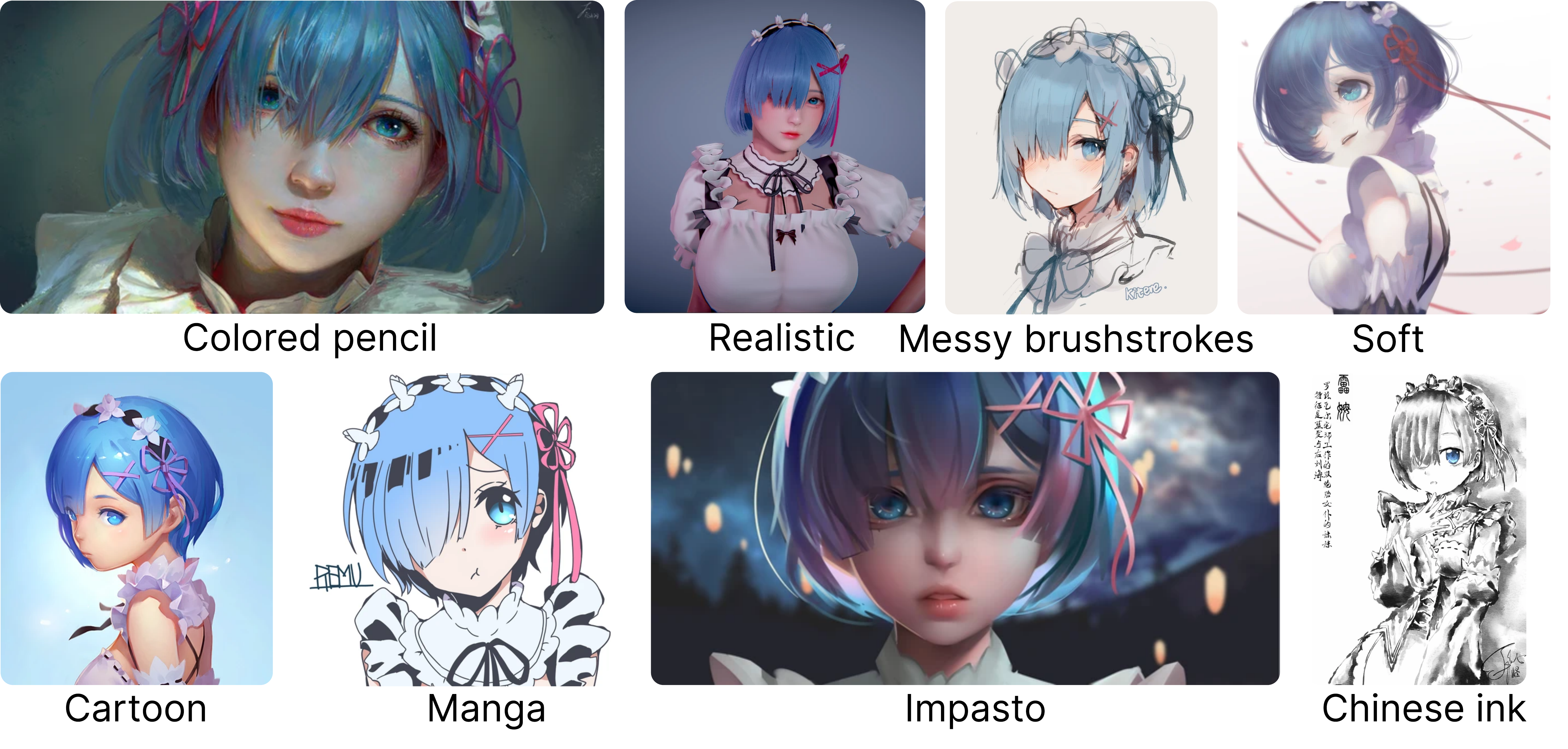}}
\subfigure[]{
\label{illust_and_label}
\includegraphics[width=0.50\textwidth]{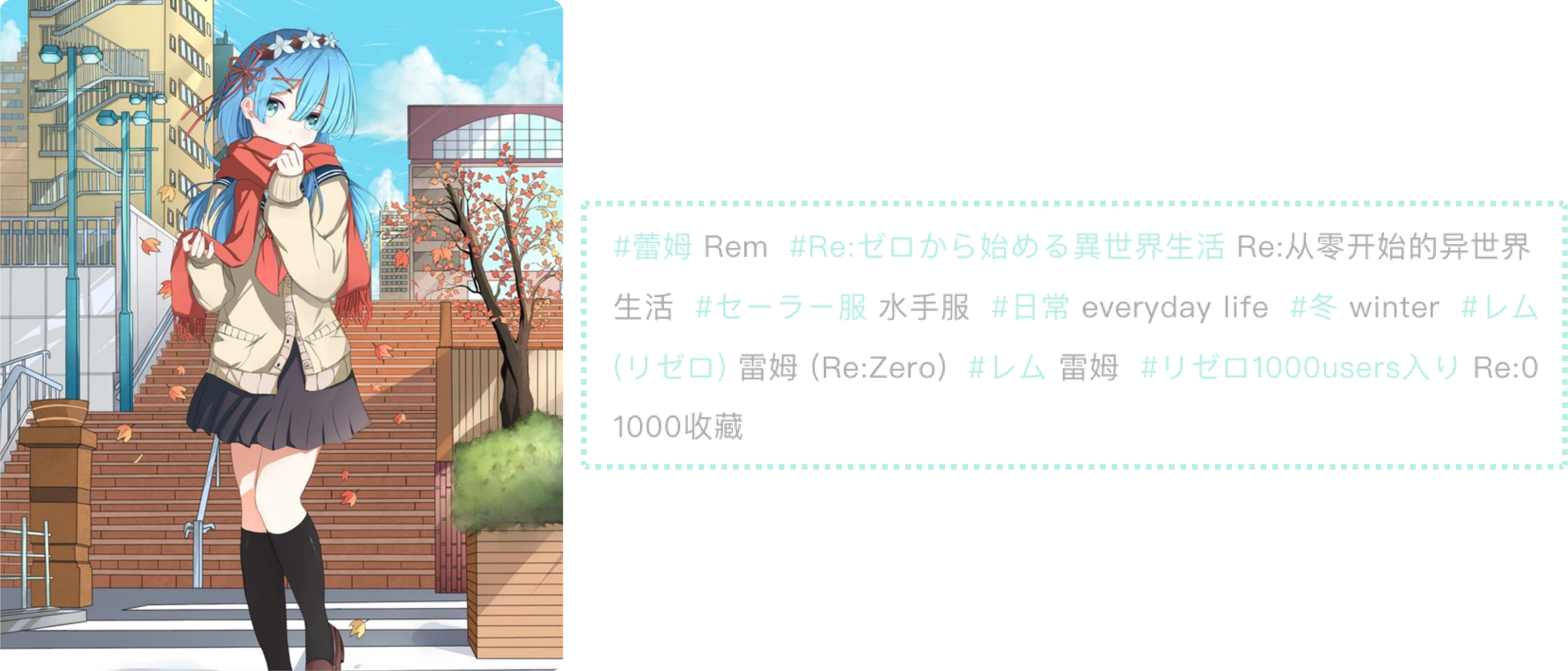}}
\caption{(a): Same content but different painting styles; (b): Image with labels.}
\label{Illustration Examples}
 \vspace{-20pt}
\end{figure}
With the booming growth and rising popularity of the ACG (Anime, Comics, and Games) industry, anime-related RS has become a research topic. AIGC models lead to an exponential increase in illustrations, and then high-quality illustrations can be generated without needing painting skills. More and more websites such as \href{https://civitai.com}{Civitai.com} are dedicated to sharing AI-generated illustrations, and users are facing the challenge of finding their favorite illustrations. To alleviate information overload, it is essential to recommend illustrations that interest users according to their preferences to improve their browsing experience. 

RS have been proven to be an effective solution for such challenges \cite{liu2023multimodal}, and many efforts have been devoted to RS. Earlier methods utilize collaborative filtering to leverage users' past interaction behaviors, including ratings and item click logs \cite{he2020lightgcn,wang2020disentangled}. Matrix factorization techniques \cite{he2017neural} and attention mechanisms \cite{zhou2018deep} have been successfully combined with deep learning to yield significant improvements. Reinforcement learning emphasizes online learning and real-time updates in models \cite{zhao2017deep}. Graph neural networks utilize information about users and items, such as social network relationships between users\cite{yang2021consisrec,yu2021self} and contextual features \cite{wang2021dcn}, to provide more personalized recommendations that align with users' interests. Recently, multi-modal recommendation systems (MRS) have gained widespread attention by leveraging different modal features of items, such as visual and textual characteristics and interaction information, to better mine item attributes not revealed in interaction.

Illustrations typically include elements such as images and text, which together form a complete story or convey information. Then MRS can help users find various media content related to their interests. However, introducing MRS to anime illustration RS still faces the following challenges: \textbf{1)} Specific feature extraction. General pre-trained encoders lack domain-specific expertise, and distinct datasets may necessitate specialized modal encoders. Nonetheless, a significant number of MRS research often resort to generic encoders or ignore the introduction of specialized encoders. \textbf{2)} Various perceptions and preferences for MM content in illustrations. E xisting MRS often fail to reweight different media modalities dynamically. The associations and interactions between multiple modalities have great potential to improve the personalization, quality, and accuracy of recommendations. It is noted that although feature crosses (FX) have been widely proven as an effective means to enhance performance in general RS, it has yet to be effectively applied in MRS.

According to the MM paradigm \cite{liu2023multimodal}, we innovated in the feature extract and MM fusion stages to address these challenges.
\textbf{1) Feature Extract. }For text encoders, in order to enable pre-training models to understand terms in the anime domain, we have constructed a large-scale anime domain multi-perspective text pair dataset that mainly includes multilingual noun mappings, relationships between entities, and explanations of nouns. Then, we fine-tune Sentence-Transformers \cite{reimers2019sentence} pre-trained models to extract text feature vectors encompassing domain knowledge. As shown in Fig. \ref{illusts_with_diff_style}, different lines, colors, brushstrokes, and composition styles can strengthen. For image encoders, we first propose simultaneously extracting both painting style and content semantic features to enhance image representation. And we construct a pretext task for multi-class multi-label prediction using images and labels as shown in Fig. \ref{illust_and_label}. Then, we construct an image encoder with dual outputs of style and semantics.

\textbf{2) MM Fusion. }We propose a User-aware Multi-modal Contribution Measurement (UMCM) mechanism that considers the various contribution levels of modalities to user preference behavior, and automatically adjusts the ratio of specific illustrations for users at the interaction level. Furthermore, considering that modalities can influence user preferences in a non-independent manner, we also introduce FX from general RS into the MM fusion stage, using DCN-V2 \cite{wang2021dcn} for higher-order modality interactions to better model user preferences.
\begin{itemize}
    \item \textbf{Insight} We emphasize the crucial importance of scene-specific modal encoders, which is generally overlooked in current MRS. As the first study on illustration MRS, we introduce and construct, for the first time, a dual-output image encoder for semantic and stylistic features. Simultaneously, we propose the concept of designing a multi-perspective domain text pair dataset to fine-tune text encoders for adaptation to specific domains.
    \item \textbf{General Framework} We analyze the varying contribution levels of multiple modalities to user preference behavior and introduce a UMCM mechanism. We also incorporate feature crosses to better model user preferences. Both of these mechanisms can be easily integrated into other MRS.
    \item \textbf{Evaluation} Our approach achieves excellent performance improvements on real-world datasets in the comparative experiments. Simultaneously, we conducted ablation experiments to explore the impact of various modules (Our codes and datasets are available on \href{https://github.com/OysterQAQ/UMAIR-FPS.git}{Github}.)
\end{itemize}

\begin{figure}
 \vspace{-10pt}
\includegraphics[width=\textwidth]{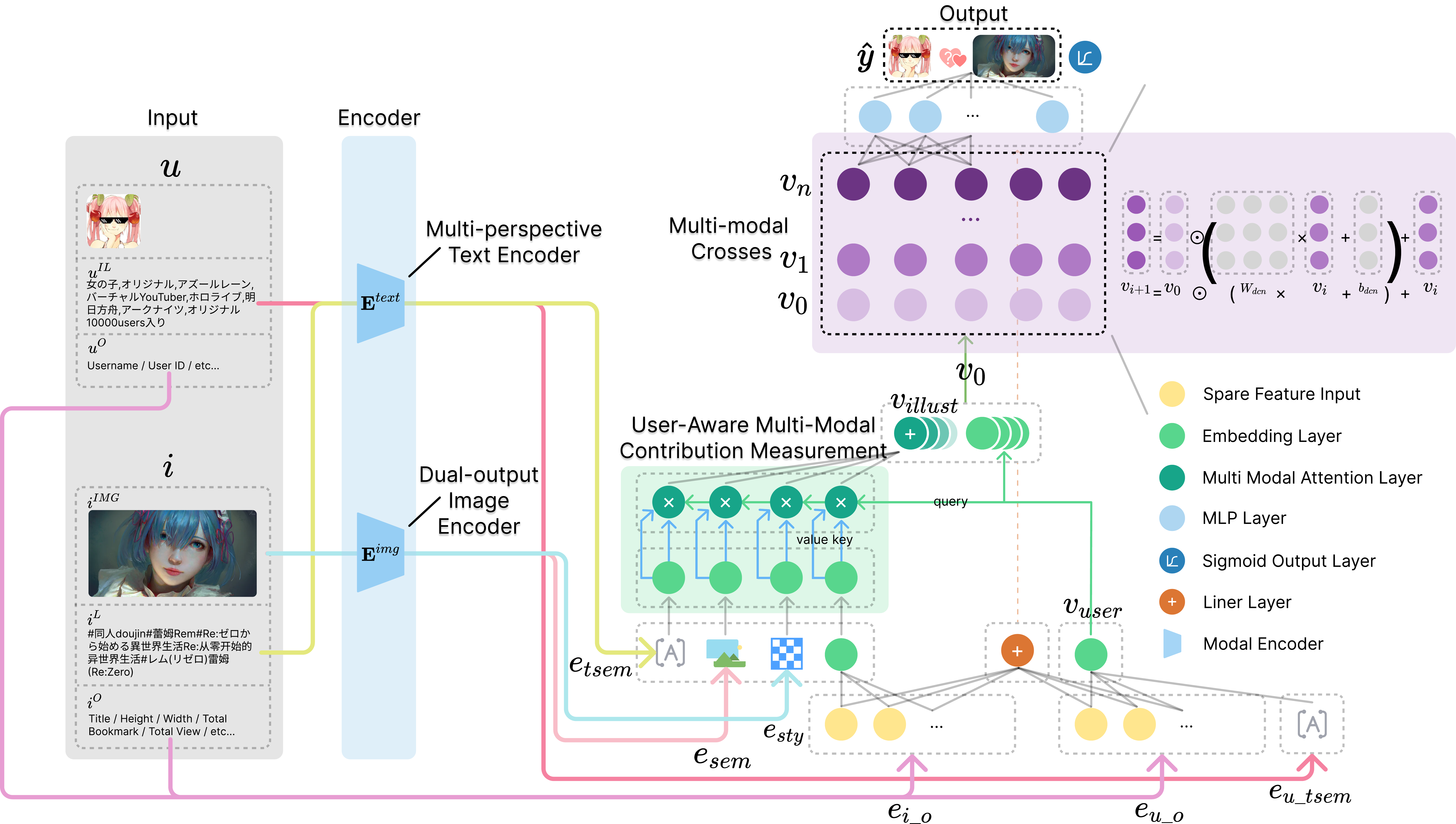}
\caption{The overview architecture of the proposed UMAIR-FPS model.}
\label{predict}
 \vspace{-10pt}
\end{figure}
\section{The Proposed Model}

\subsection{Problem Definition}


The input of the anime illustration RS task is a set of $N$ users $u$ as $\mathcal{U}=\left\{u_1,  \ldots, u_{N}\right\}$ and a set of $M$ illustrations as
 $\mathcal{I}=\left\{i_1, \ldots, i_{M}\right\}$. Specifically, user $u=(u^\text{IL}, u^\text{O})$, where $u^\text{IL}$ represents interest label, $u^\text{O}$ is profile that contains username, gender, etc. Illustration $i=(i^\text{IMG}, i^\text{L},i^\text{O})$, where $i^\text{IMG}$ is image of $i$, $i^\text{L}$ represents labels that artists use to annotate $i$, and $i^\text{O}$ implies other metadata of illustration such as publish date, image size, etc. And the $u$-$i$ interaction is typically formulated as a matrix $Y=\{y\}_{N\times M}$, specifically, $y = 1$ means $u$ has bookmarked $i$, otherwise $y = 0$. Now, our learning goal is to train a model to predict the probability $\hat y$ of $u$ clicks the target $i$, formulated as:
\begin{equation}
\begin{aligned}
 \hat y=\mathcal{F}(\theta,u,i),
 \label{eq:pd1}
\end{aligned}
\end{equation}
where $\mathcal{F}$ and $\theta$ denote the model and its weights. 

\subsection{Dual-output Image Encoder}
\label{dual-optput}

\begin{figure}
 \vspace{-20pt}
\centering  
\subfigure[]{
\label{image_encoder}
\includegraphics[width=0.45\textwidth]{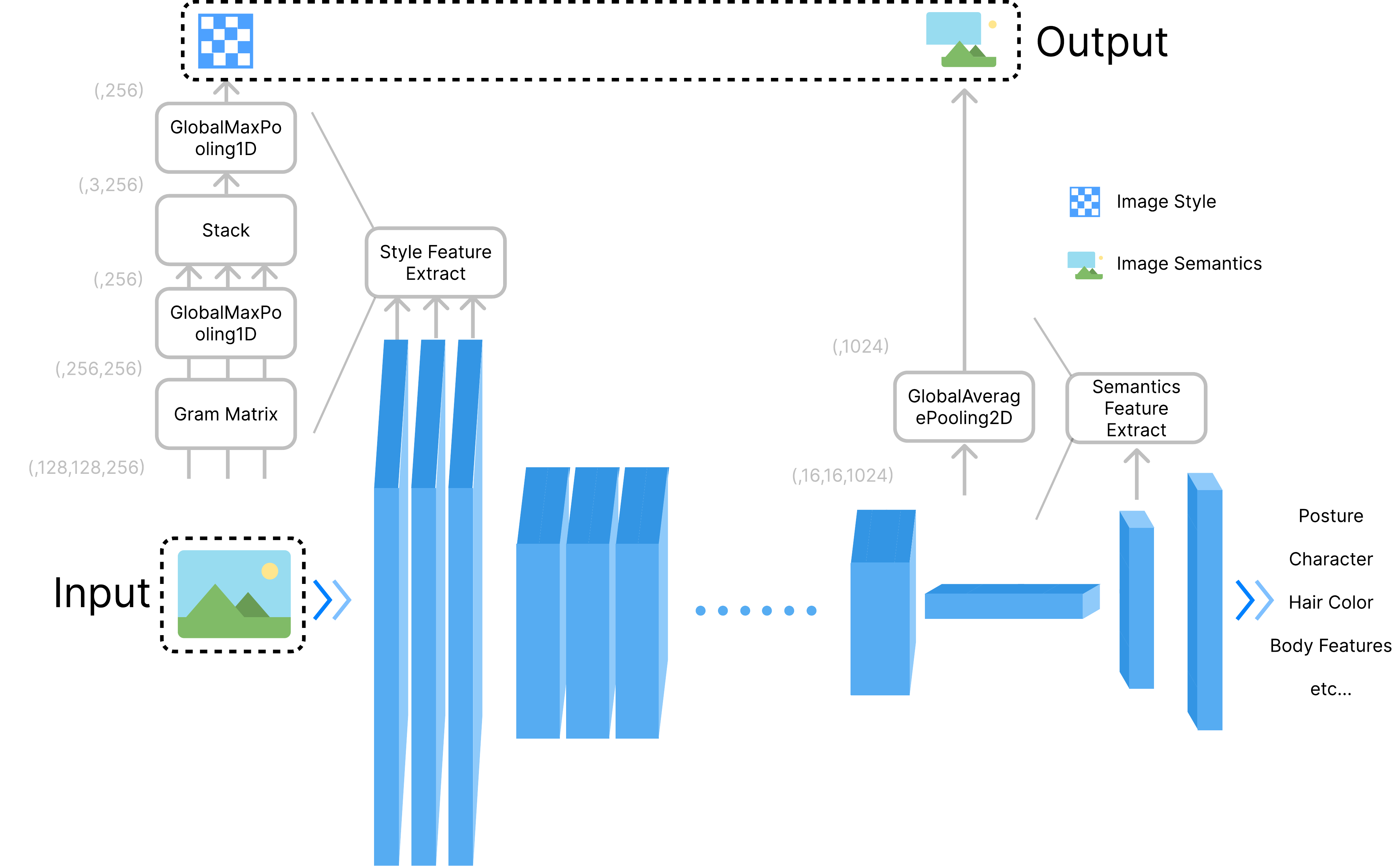}}\hspace{4mm}
\subfigure[]{
\label{text_encoder}
\includegraphics[width=0.45\textwidth]{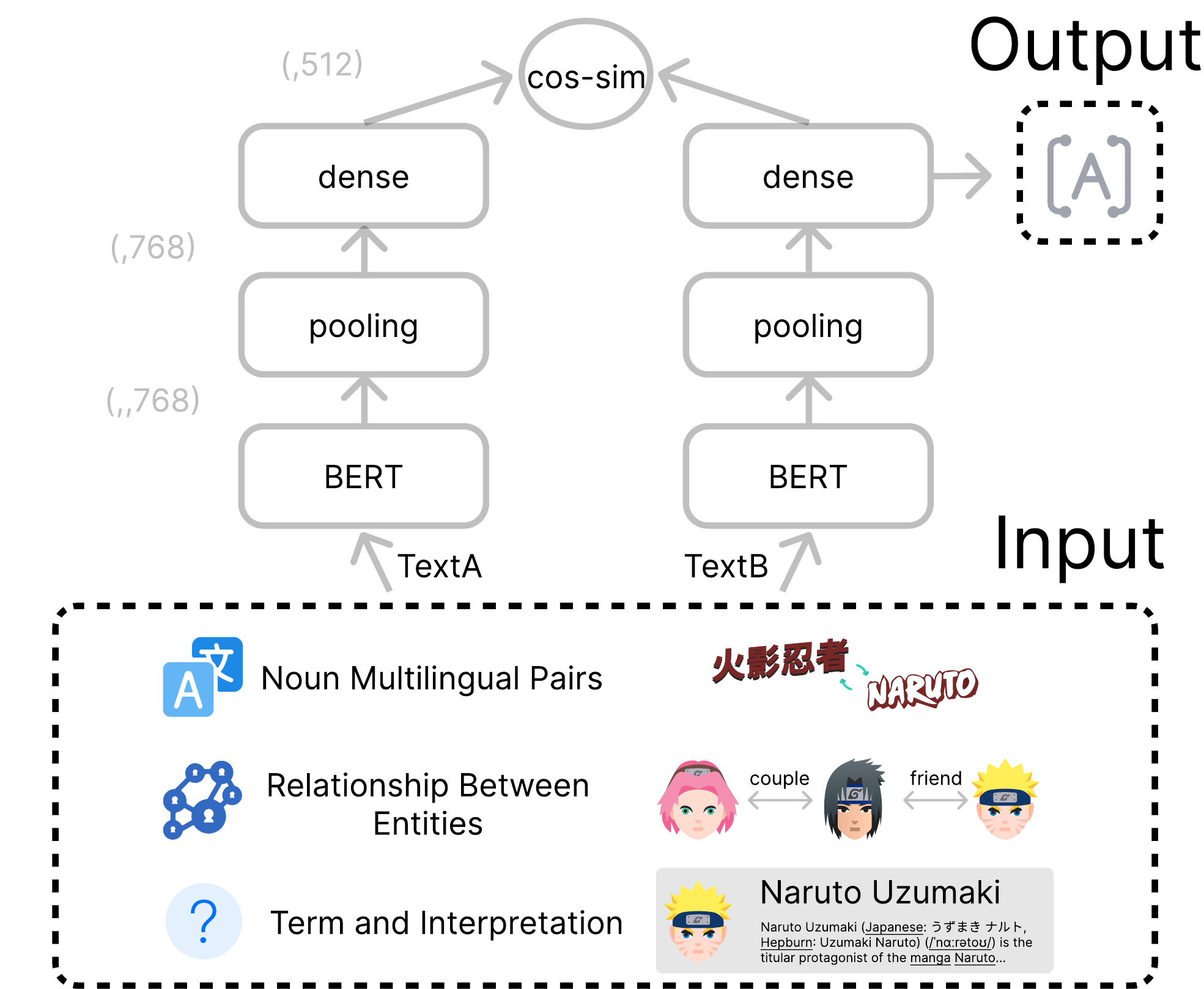}}
\caption{Dual-output image encoder \& Multi-perspective text encoder.}
 \vspace{-10pt}
\end{figure}

Different painting styles give the illustration personality and artistry, making it more attractive, expressive, and recognizable. In the field of illustrations, not only the semantic features of its images but also the stylistic features of its paintings influence user preferences. Therefore, we have built an image encoder $\mathbf{E}^\text{img}$ with the semantic feature $e^\text{SEM}$ and style feature $e^\text{STY}$ outputs according to the following steps: \textbf{1)} To extract image features that align with animation domain knowledge, we use $i^\text{IMG}$ as features and $i^\text{L}$ as labels, building a large-scale anime image multi-label multi-classification dataset. \textbf{2)} We chose the structure of ResNet101 to construct the encoder and pre-train on the dataset. 3) The classification head is removed from the pre-training model after fitting the pretext task. Then, as shown in Fig. \ref{image_encoder}, the feature map (FM) generated from various layers of $\mathbf{E}^\text{img}$ is modified to obtain dual outputs: 

\begin{equation}
\begin{aligned}
e^\text{SEM}, e^\text{STY}=\mathbf{E}^\text{img}(i^\text{IMG}).
 \label{eq:doie1}
\end{aligned}
\end{equation}


\noindent\textbf{Semantic Feature Output. } Higher-level FM is obtained from convolutional neural networks (CNN) to describe more complex features. In the pretext task, we chose the text annotations $i^\text{L}$ made by artists as the labels for the images. As a result, $\mathbf{E}^\text{img}$ can effectively learn these semantic features represented by pre-trained task. For reducing the spatial resolution of features to capture high-level semantic features of images better, we add a mean pooling layer after the output of the last convolutional layer of $\mathbf{E}^\text{img}$ to obtain feature $e^\text{SEM}$ as:

\begin{equation}
\begin{aligned}
e^\text{SEM} = \text{MeanPooling}(\mathbf{E}^\text{img}_{last}(i^\text{IMG})),
 \label{eq:doie2}
\end{aligned}
\end{equation}
where $\mathbf{E}^\text{img}_{last}(i^\text{IMG})$ represents the outputs of the last convolutional layer of $\mathbf{E}^\text{img}$.

\noindent\textbf{Style Feature Output. }As each layer's FM contains multiple channels, using shallow FMs directly as style features not only leads to excessive output dimensions but also fails to capture the interactions between channels. Since these interactions are closely related to the texture and style of the image, we adopt the channel-wise Gram matrix to capture the interactions between channels in the FMs and quantify style features as follows:
\begin{equation}
\begin{aligned}
g^l_{j,k} = \frac{\sum_{h=0}^{H} \sum_{w=1}^{W} \mathbf{E}^\text{img}_{l,j,h,w}(i^\text{IMG})\mathbf{E}^\text{img}_{l,j,h,w}(i^\text{IMG})}{H \times W},
 \label{eq:doie3}
\end{aligned}
\end{equation}
where $\mathbf{E}^\text{img}_{l,j,h,w}(i^\text{IMG})$ represents the pixel value at position $(h, w)$ on channel $j$ in $l$-th layer of the $\mathbf{E}^\text{img}$, $\mathbf{E}^\text{img}_{l,k,h,w}(i^\text{IMG})$ represents same for channel $k$. The size of the FM in the specific convolutional layer is consistent, and its height and width are represented by $H$ and $W$.

Since direct combination will result in excessive output dimensions, max pooling is utilized to remove local details of the image and allows the FM to focus more on the global texture and structure of the image. This abstract representation is more suitable for representing style feature of an image. Therefore, the layer-wise Gram matrix representation and $e^\text{STY}$ are given by merging the $G^l$ of the first three layers by max pooling.


\subsection{Multi-perspective Text Encoder}

Although general pre-trained models (GPMs) perform well in various natural language processing tasks, they often face challenges when adapting to specific domains. For example, misunderstandings or errors may occur in domain-specific terminology and knowledge. It is difficult for GPMs to understand vocabulary such as character names, specific anime titles, concepts, and settings within animation works. The combination of multilingualism such as Chinese, Japanese, and English and domain-specific vocabulary leads to poor performance of GPMs in our datasets. Therefore, we obtain multi-perspective text pair to fine-tune the GPM and construct a text encoder $\mathbf{E}^\text{text}$ integrated with domain knowledge from diverse views as two main steps: 


\textbf{1)} As shown in Fig. \ref{text_encoder}, we collect publicly available information from various websites such as \href{https://bangumi.tv}{Bangumi}, \href{https://zh.moegirl.org.cn}{Moegirlpedia}, \href{https://myanimelist.net}{MyAnimeList}, \href{https://www.wikipedia.org}{Wikipedia} etc., to create a dataset consisting of text pairs from multiingual relation, domain relationship, term explanation perspectives as shown in Table \ref{multi-perspective-text}.
\begin{table}[]
  \vspace{-20pt}
  \caption{Perspective types of text pairs.}
\centering
\scriptsize
	\begin{tabular}{c|cc}
		\hline
		Type &
		\multicolumn{2}{c}{Sub Type \& Example} \\ \hline
		\multirow{3}{*}{\begin{tabular}[c]{@{}c@{}}\textbf{Multilingual}\\ mappings of\\ domain terms\end{tabular}} &
		\multicolumn{1}{c|}{Chinese-English} &
		\begin{CJK}{UTF8}{gbsn}火影忍者\end{CJK} $\boldsymbol{\leftrightarrow}$ \textit{Naruto}\\ \cline{2-3} 
		&
		\multicolumn{1}{c|}{English-Japanese} &
		\textit{Uchiha Sasuke} $\boldsymbol{\leftrightarrow}$ \begin{CJK}{UTF8}{gbsn}うちはサスケ\end{CJK} \\ \cline{2-3} 
		&
		\multicolumn{1}{c|}{Japanese-Chinese} &
		 \begin{CJK}{UTF8}{gbsn}うずまきナルト\end{CJK} $\boldsymbol{\leftrightarrow}$ \begin{CJK}{UTF8}{gbsn}漩涡鸣人\end{CJK} \\ \hline
		\multirow{2}{*}{\begin{tabular}[c]{@{}c@{}}\textbf{Relationships}\\ Between domain entities\end{tabular}} &
		\multicolumn{1}{c|}{series-character} &
	\textit{Naruto}'s character $\boldsymbol{\leftrightarrow}$ \textit{Uchiha Sasuke}\\ \cline{2-3} 
		&
		\multicolumn{1}{c|}{character-character} &
		\textit{Uchiha Sasuke}'s friend $\boldsymbol{\leftrightarrow}$ \textit{Uzumaki Naruto} \\ \hline
		\multirow{2}{*}{\begin{tabular}[c]{@{}c@{}}\textbf{Explanations}\\ of\\ domain terms\end{tabular}} &
		\multicolumn{1}{c|}{\begin{tabular}[c]{@{}c@{}}animation name and\\ its introduction\end{tabular}} &
		\begin{tabular}[c]{@{}c@{}}\textit{Naruto}  $\boldsymbol{\leftrightarrow}$ \textit{Naruto} is a Japanese manga series\\ that tells the story of \textit{Naruto Uzumaki}...\end{tabular} \\ \cline{2-3} 
		&
		\multicolumn{1}{c|}{\begin{tabular}[c]{@{}c@{}}animation setting and \\ its meaning\end{tabular}} &
		\begin{tabular}[c]{@{}c@{}}\textit{chakra} $\boldsymbol{\leftrightarrow}$ \textit{chakra} is a Sanskrit word that\\ means wheel or cycle.\end{tabular} \\ \hline
	\end{tabular}
  \label{multi-perspective-text}
   \vspace{-10pt}
\end{table}

Specifically, texts from different sources can provide multi-hierarchy information about a specific topic. The comparison of multiple languages enables the model to learn the correspondence between multiple languages.

\textbf{2)} We fine-tune the Sentence-Transformers \cite{reimers2019sentence} with multi-language pre-trained weight for semantic search on the dataset. And a text encoder $\mathbf{E}^\text{text}$ is obtained as integrating domain knowledge to output textual semantic features as:
\begin{equation}
\begin{aligned}
e^\text{TSEM}=\mathbf{E}^\text{text}(i^\text{L}).
 \label{eq:mte2}
\end{aligned}
\end{equation}

\subsection{User-aware Multi-modal Contribution Measurement}

We obtain a list of MM representations $e^\text{MM}= \{e^\text{TSEM},e^\text{SEM},  e^\text{STY},  F^\text{emb}(i^\text{O})\}$, where the vector representation of metadata $i^\text{O}$ is obtained through an embedding layer $F^\text{emb}$. Representations from different modes should have varying levels of contribution to user preference. For example, a user $u$ might bookmark $i$ with diverse preferences due to stylistic features, semantic content of illustrations or text labels. Guided by interaction data, we propose the UMCM mechanism to learn the dynamic weight of the MM features of illustration, and give the re-weighted MM illustration features $v^\text{il}$ as:
\begin{equation}
\begin{aligned}
v^\text{il} = \sum_{j=1}^J \alpha(v^\text{user}, e^\text{MM}_j)e^\text{MM}_j,
 \label{eq:umcm2}
\end{aligned}
\end{equation}
where $v^\text{user}$ refers to the user $u$ feature vector obtained by embedding $u^\text{IL}$ and $u^\text{O}$, function $\alpha(\cdot)$ outputs the weights of MM features.  The function $\alpha(\cdot)$, implemented by the  dot-product attention mechanism, is defined as:

\begin{equation}
\begin{aligned}
\alpha(v^\text{user}, e^\text{MM}_j) = \frac{\exp \left(a_j\right)}{\sum_{j=1}^J \exp \left(a^j\right)},\ {\text{where}} \ a_j=v^\text{user} \odot e^\text{MM}_j.
 \label{eq:umcm3}
\end{aligned}
\end{equation}

Given the effectiveness of individual UMCM, inspired by the MoE \cite{ma2018modeling} framework in multi-task RS, we employ parallel UMCMs to simulate experts for different decisions. Furthermore, we utilize the dot-product attention mechanism to implement an aggregation gate, consolidating the decisions from multiple experts, thereby further enhancing the modeling of MM contributions.

\subsection{Multi-modal Crosses}

Utilizing MLP to simulate implicit feature intersections of any order is time-consuming. To avoid unnecessary computations, we utilize a lightweight DCN-V2 \cite{wang2021dcn} module to simulate explicit and bounded-degree MM crosses. The core idea of DCN-V2 is to create explicit feature cross layers. 



\section{Experiments}



\subsection{Experimental Settings}

\begin{figure}
 \vspace{-30pt}
\centering  
\subfigure[]{
\label{fig:user_interaction_freq}
\includegraphics[width=0.45\textwidth]{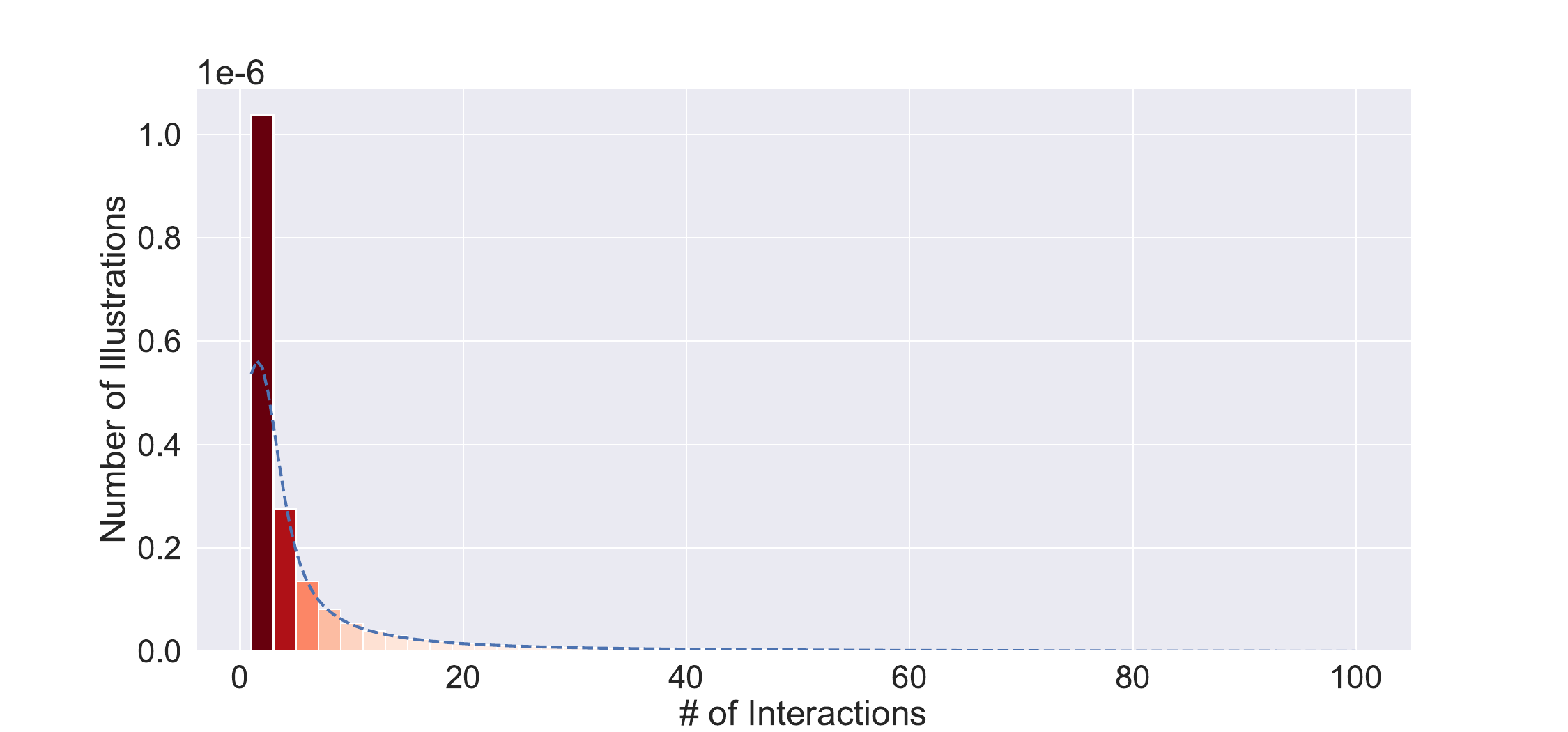}}
\subfigure[]{
\label{fig:illust_interaction_freq}

\includegraphics[width=0.45\textwidth]{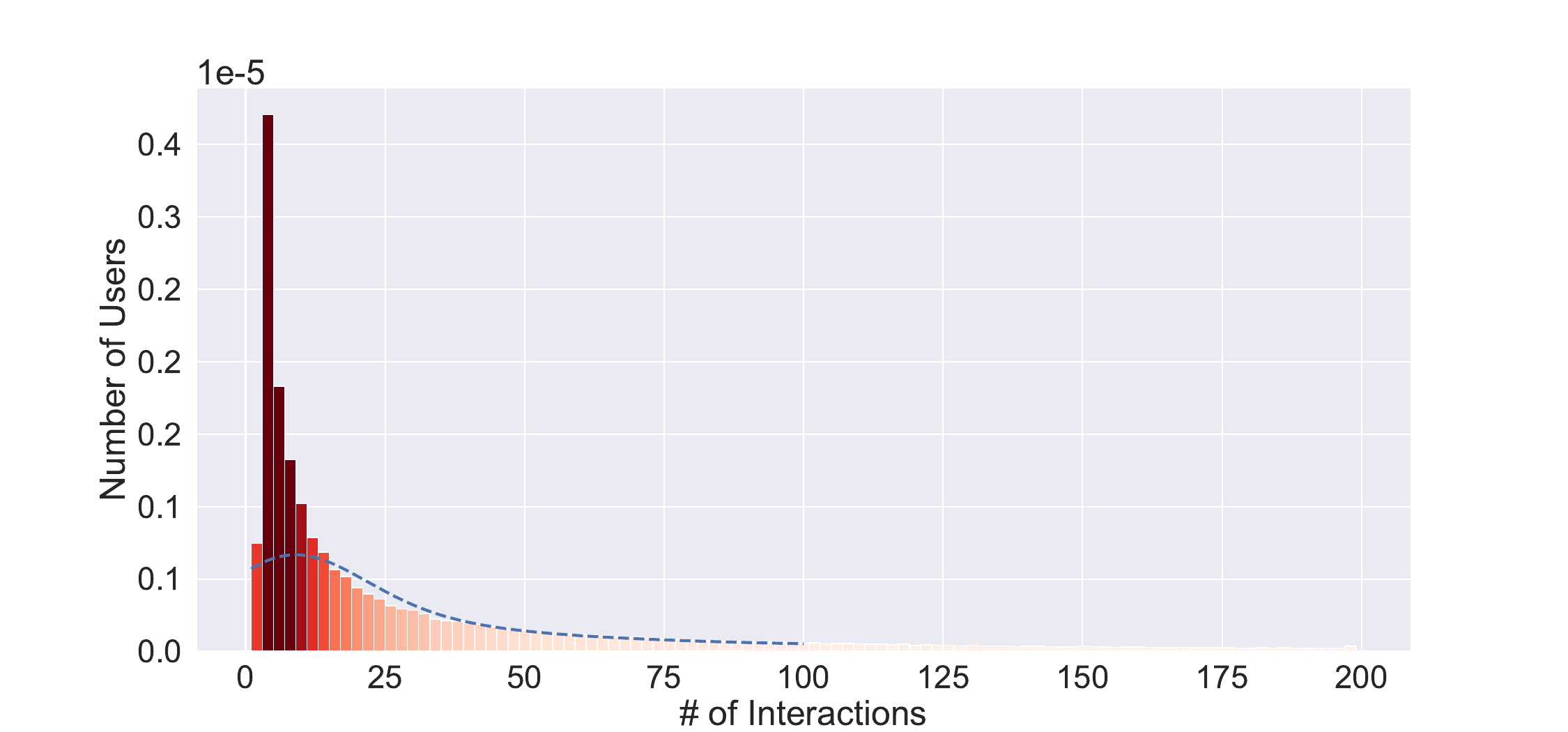}}
\caption{(a)/(b): Interaction-based Illustration/User statistics.}
 \vspace{-10pt}
\label{fig:userandillustraion}
\end{figure}
\begin{table}
\vspace{-35pt}
\caption{ Detailed information of the dataset}
\centering
\begin{tabular}{c|c|c|c|c|c}
\hline
\#Dataset & \#Category & \#User  & \#Illustrations & \#Interactions & Time range            \\ \hline
$\mathcal{D}$  &   Train    & 215,394 & 1,882,675       &24,233,663      & 2021/01/01-2021/10/01 \\ \hline
$\mathcal{D}_t$  &  Test  & 108,955 & 1,229,347       & 9,472,225      & 2021/10/01-2022/01/01 \\ \hline
\end{tabular}

\label{tab:detail}
\vspace{-10pt}
\end{table}
\textbf{Dataset. }To the best of our knowledge, there currently exists yet to be a high-quality public dataset for anime illustration RS. We collected user interaction logs from a commercial illustration website to construct our dataset. The detailed information of the dataset is shown in Table \ref{tab:detail}. Fig. \ref{fig:userandillustraion} (a) and (b) depict the distribution of interaction counts from the perspectives of illustrations and users, respectively, showcasing the long-tail distribution challenge present in the dataset, underscoring the complexities involved in anime illustrations RS.

\textbf{Evaluation Metrics.} As in \cite{lian2018xdeepfm,lu2021dual,huang2019fibinet,wang2021dcn}, we use the Area Under the Curve and Binary Cross-Entropy Loss to evaluate the performance of all methods.

\textbf{Baselines.} 
To demonstrate the effectiveness of our proposed method, we compare it with six SOTA general RS approaches. 



\textbf{Implementation Details.} We implement UMAIR-FPS on the TensorFlow framework, and use the Adam optimizer with a learning rate of 1e-3 and train on the dataset for just one epoch to avoid the overfitting issue \cite{zhang2022towards}. We adhere to default settings of most hyperparameters in the baseline methods except outputting an embedding layer with a dimension of 8 and an L2 regularization parameter of 1e-3.

\subsection{Performance Comparison}

\begin{table}
    \vspace{-20pt}
    \caption{Performance comparison on $\mathcal{D}_t$. Bold: best. Underline: runner-up.}
    \centering
       \begin{tabular}{l|c|c|c}
    \hline
       Metric & BCE Loss & AUC & Parameter \\ \hline
        DCN\cite{wang2017deep}  & 1.1964 & 0.7766 & \underline{10,175,324} \\ 
        xDeepFM\cite{lian2018xdeepfm}  & 1.7769 & 0.7219 & 10,302,596 \\ 
        FiBiNET\cite{huang2019fibinet}  & 1.3760 & 0.7019 & 10,424,966 \\ 
        DIFM\cite{lu2021dual}  & 0.5685 & \underline{0.8055} & 10,210,660 \\ 
        DCN-V2\cite{wang2021dcn}   & 0.6469 & 0.7726 & 10,278,092\\ 
        FinalMLP\cite{mao2023finalmlp}  & \underline{0.5087}  & 0.8034  & \boldmath{${10,160,912}^{*}$} \\ 
        \hline
        \textbf{UMAIR-FPS} & \boldmath{${0.3797}^{*}$} & \boldmath{${0.8490}^{*}$} & 10,597,028 \\ 
        
        \textbf{Improvement} & \textbf{25.35\%} & \textbf{5.4\%} & -4.292\% \\ \hline
    \end{tabular}
\label{baseline}
\vspace{-10pt}

\end{table}

Table \ref{baseline} presents the performance comparison of UMAIR-FPS with all other methods on the dataset. It also reports the parameters for each model to assess the space complexity. As shown in Table \ref{baseline}, our model exhibits significant improvement compared to six baseline models. Compared to the best-performing baseline, UMAIR-FPS still improves AUC by 5.4\% and, notably, boosts the BCE loss by an impressive 25.35\%. We believe the superior performance of our model can be attributed to the following factors: \textbf{1) Combination of textual and visual information.} This allows us to model the intrinsic correlations, generating a more comprehensive item representation, and to some extent, alleviates data sparsity caused by the long-tail distribution. \textbf{2) UMCM mechanism.} By re-weighting the MM features, our model can simulate the real selection process of users more closely. \textbf{3) MM crosses.} This increases the non-linearity capabilities of the model, capturing user preferences in a more granular manner.

Even though UMAIR-FPS introduces features from multiple modalities, its parameter count remains on par with other models. This ensures that our method does not compromise efficiency while guaranteeing predictive accuracy.

\subsection{Ablation Study}
To further investigate the effectiveness of our key modules, we conduct ablation studies on four groups of variants, as presented in Table \ref{ablation}.

\begin{table}
 \vspace{-20pt}
\tiny
\caption{Ablation study on key components of UMAIR-FPS.}
    \centering
 \begin{tabular}{@{}c|c|c|c|c|c|c|c|c|c@{}}
  \hline
     Group  &  Variants & BCE Loss & AUC & Parameter&Group  &  Variants & BCE Loss & AUC & Parameter \\ \hline
  \multirow{4}{*}{Multi-modal} &  ${w/o\ \text{ALL}}$  &  0.6469 & 0.7726 & 10,283,072 &\multirow{6}{*}[-0.5em]{\makecell{UMCM\\ by MoE}} &       ${w/\ \text{ATT*2}}$ &  0.5846 & 0.8341 & 10,597,028 \\
     &   ${w/o\ \text{STY}}$ & 0.5267 & 0.8240 & 10,580,580 & &    ${w/\ \text{ATT*3}}$&  0.3797 &  0.8490 & 10,597,028 \\ 
     &    ${w/o\ \text{SEM}}$  &  0.5201 & 0.8433 & 10,547,812 &&   ${w/\ \text{ATT*4}}$  & 0.4214  & 0.8475 & 10,597,028 \\ 
     & ${w/o\ \text{TSEM}}$  & 0.5943  & 0.8233  & 10,564,196 &&       ${w/\ \text{SEN*2}}$ & 0.5123 & 0.8370 & 10,598,188 \\
\cline{1-5}
      \multirow{3}{*}[-0.1em]{\makecell{UMCM by\\ Single}} &  ${w/o\ \text{UMCM}}$ & 0.5493 & 0.8297 & 10,556,624 &&    ${w/\ \text{SEN*3}}$  & 0.4963 & 0.8465 & 10,598,768 \\ 
     &   ${w/\ \text{UMCM\_ATT}}$ & 0.4750 & 0.8470 & 10,597,028 &&     ${w/\ \text{SEN*4}}$   & 1.6368 & 0.7607 & 10,599,348 \\  
\cline{6-10}
     &  ${w/\ \text{UMCM\_SEN}}$  &0.5172 & 0.8484 & 10,597,608 & {DCN-V2}  &   ${w/o\ \text{DCN-V2}}$ & 0.5777 & 0.8337 & 10,522,532  \\ 
\hline   
- & UMAIR-FPS & 0.3797 & 0.8490 & 10,597,028&-&-&-&-&-\\ 
\hline
    \end{tabular}
\label{ablation}
\vspace{-10pt}
\end{table}

\textbf{ Impact of Multi-modal Features.}
To explore the effectiveness of the MM features, we establish four variants. In Variant ${w_/o\ \text{ALL}}$, we eliminate all MM features, including those representing the style and semantic features of images, as well as semantic features of text labels. As a result, the AUC decreases by 9.00\%, the BLE loss increases by 70.37\%, and the parameter count drops by 2.96\%. This trend underscores the significance of combined text and visual multimodal features in accurately capturing user preferences, and the model's performance is hardly impacted by this integration.

In Variants ${w_/o\ \text{STY}}$, ${w_/o\ \text{SEM}}$, and ${w_/o\ \text{TSEM}}$, we respectively remove the style features of images $e^\text{STY}$, the semantic features of images \(e^{\text{SEM}}\), and the semantic features of text labels \(e^{\text{TSEM}}\). The most notable impact comes from removing $e^\text{STY}$, indicating that it bring substantial improvements in anime recommendations. Among all features, the enhancement brought by \(e^{\text{SEM}}\) is the smallest, as it is guided by the actual image text labels. Moreover, the information obtained by selecting lower-level features might still overlap with \(e^{\text{TSEM}}\).

\begin{figure}
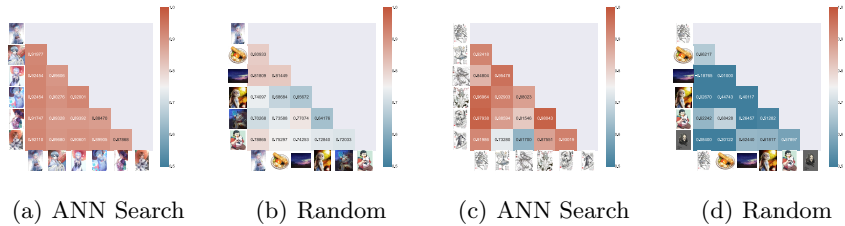

 \vspace{-23pt}
\centering  
\subfigure[ANN Search]{
\label{semantics_search}
\includegraphics[width=0.225\textwidth]{image/semantics_search.pdf}}
\subfigure[Random]{
\label{semantics_random}
\includegraphics[width=0.225\textwidth]{image/semantics_random.pdf}}
\subfigure[ANN Search]{
\label{style_search}
\includegraphics[width=0.225\textwidth]{image/style_search.pdf}}
\subfigure[Random]{
\label{style_random}
\includegraphics[width=0.225\textwidth]{image/style_random.pdf}}
\caption{Compare $e^\text{SEM}$ between (a) \& (b), and compare $e^\text{STY}$ between (c) \& (d).}
\label{heatmap}
 \vspace{-15pt}
\end{figure}

The effectiveness of the image encoder $\mathbf{E}^\text{img}$ is shown in Fig. \ref{heatmap}. The cosine distances between the feature vectors of similar artworks at the semantic feature and the style feature level are close, while they are far for dissimilar ones.

\textbf{Impact of UMCM. } 
We design two new variants: one without UMCM, denoted as ${w_/o\ \text{UMCM}}$, one replaced the ATT with the SENet module, denoted as ${w/\ \text{UMCM\_SEN}}$, and one implemented with the dot product attention mechanism, ${w/\ \text{UMCM\_ATT}}$. Compared to ${w_/o\ \text{UMCM}}$, ${w/\ \text{UMCM\_ATT}}$ reduces the loss by 13.52\%, increases the AUC by 2.085\%, and augments the model parameters by 0.382\%. Compared to ${w/\ \text{UMCM\_SEN}}$, the loss decreases, while the AUC remains roughly consistent. This suggests that dynamically reweighting illustrations based on user feature vectors can effectively model user preferences.

When utilizing UMCM by MoE, We enhance UMCM with the MoE framework. Our observations indicate that, for multimodal contribution reweighting, a three-module stack is optimal for both ATT and SNET. Adding additional modules tends to degrade performance. Thus, after introducing MM cross-interactions into the multi-task RS framework, we can enhance recommendation performance by leveraging multiple weight modeling modules formed by expert outputs, facilitating representation learning for specific tasks.

\textbf{Impact of MM Crosses.} We compare UMAIR-FPS with the variant without FX, denoted as ${w_/o\ \text{DCN-V2}}$. denoted as ${w_/o\ \text{DCN-V2}}$. The AUC decreases by 1.80\% and the BCE loss increases by 52.14\%, with the model parameters reduced by 0.70\%. These results underscore the efficacy of  FX. Moreover, undergoing user-aware weight fusion, can still benefit from FX with user features.








\section{Conclusion}
This paper introduces UMAIR-FPS, a novel user-aware MM anime illustration RS. For the first time, we propose integrating painting style into MM features. Our method constructs a dual-output image encoder based on the stylistic and semantic features and a text encoder that is fine-tuned using multi-perspective text, enabling it to understand anime knowledge. Additionally, to account for the varying contribution levels of multiple modalities to user preference behavior, we design UMCM based on attention. Conducting FX among multiple modalities utilizing DCN-V2 further enhances the model's recommendation accuracy. Through extensive experiments on real-world datasets, we validate the superiority of UMAIR-FPS compared to other SOTA methods.

%
%
%
%
\bibliographystyle{splncs04_unsort}
\bibliography{reference}
\end{document}